\documentclass[10pt]{iopart}
\usepackage{graphicx,hyperref,amsfonts,amssymb,mathrsfs,comment}
\usepackage{epstopdf,color,bm,multirow,rotating,soul,comment}
\usepackage[absolute,overlay]{textpos}
\usepackage{graphicx}
\usepackage{subfigure}
\usepackage{graphicx,hyperref}
\expandafter\let\csname equation*\endcsname\relax
\expandafter\let\csname endequation*\endcsname\relax
\usepackage{amsmath}
\usepackage{amssymb}
\usepackage{multirow, makecell, comment} 
\newcommand{\fla}[1]{\begin{flalign}#1\end{flalign}}
\usepackage{braket}
\usepackage{lineno}
\usepackage{lipsum}
\usepackage{xcolor}
\usepackage{soul}
\usepackage{array}

\begin{document}

\title[Integrated multiresonator quantum memory]{Integrated multiresonator quantum memory}
\author{N S  Perminov$^{1,2}$, and S A Moiseev$^{1,2,*}$}

\address{$^{1}$ Kazan Quantum Center, Kazan National Research Technical University n.a. A.N.Tupolev-KAI, 10 K. Marx, Kazan 420111, Russia}
\address{$^{2}$ Zavoisky Physical-Technical Institute, Kazan Scientific Center of the Russian Academy of Sciences, 10/7 Sibirsky Tract, Kazan 420029, Russia}
\ead{$^{*}$s.a.moiseev@kazanqc.org}

\vspace{10pt}
\begin{indented}
\item[] \today
\end{indented}

\begin{abstract}
We develop the integrated efficient multiresonator quantum memory scheme based on a system of three interacting resonators coupled through a common resonator to an external waveguide via switchable coupler. It is shown that high-precision parameter matching based on the step-by-step optimization makes it possible to efficiently store the signal field and on demand retrieval the signal at specified time moments. Possible experimental implementations and practical applications of the proposed quantum memory scheme are discussed.
\end{abstract}

\vspace{2pc}
\maketitle
\ioptwocol

\section{Introduction}
Integrated optical quantum memory (QM) is of great importance for the successful development of quantum information technologies \cite{Moody_2022,Blais2020}.
There are many promising approaches in the elaboration of efficient QM intensively discussed in the reviews \cite{Hammerer2010,Lvovsky2009,Tittel2010,Khabat2016,Chaneliere2018}.
Among the approaches being developed, cavity assistant QM schemes are attracting increasing attention, which can significantly enhance the interaction of photons with emitters placed in the resonator 
\cite{Cirac_1997,Fleischhauer_2000,Gorshkov_2007,Boozer_2007,Moiseev2010,Afzelius2010,Dilley_2012,Reiserer_2015,Giannelli_2018,Kaczmarek_2018,Veselkova_2019,Arslanov_2021}.
The active development of high-quality resonators and the development of integrated technologies made it possible to create multi-resonator systems that significantly facilitate the control  \cite{Li_2017} and  generation  of light fields \cite{Chuprina_2018}, that made these systems also promising for use in quantum memory \cite{McKay2015,2017-Naik-NatureComm,Moiseev_2017_PRA,
EMoiseev2017,Moiseev2018,Bao2021,Matanin2022}.
In these works it was shown that this advantage is due to the fact that the multi-resonator schemes easily provides efficient transfer of broadband input signal pulse from external waveguide to the systems of high-Q resonators which can be further used for the long-lived storage and quantum processing. Below we are developing a multi-resonator QM \cite{Moiseev_2017_PRA,EMoiseev2017,Moiseev2018,Bao2021,Matanin2022} 
in which the resonators have different frequencies and can cover an arbitrarily wide spectral range, which makes this approach particularly promising for the implementation of broadband QM.

In addition, the system of high-Q resonators with controlled frequencies allows implementing on demand broadband QM \cite{Moiseev_2017_PRA}, which was demonstrated experimentally on a chip of four superconducting resonators \cite{Bao2021} for  microwave pulses attenuated to a single-photon level. 
The realized four-resonator QM have demonstrated the ability for storage of two pulses and rich functionality for the implementing various controlled dynamics in transformation of quantum information and switching between different storage modes due to the ability to quickly changing the frequencies of resonators.
However experimentally achieved efficiency 6\% in \cite{Bao2021} is still not high enough, which indicates the need to develop more effective ways to control the interaction of the resonator system with an external waveguide.

Recently, we have found \cite{Perminov_2019} that is possible to implement even highly efficient QM containing a countable small number of resonators which require precise coordination of a number of its physical parameters and geometric scale.
Efficient quantum storage can be easily realized for multi-resonator QM with one common resonator connected to an external waveguide \cite{Moiseev2018,Matanin2022}.
Superefficient broadband quantum storage (with efficiency $\sim 0.999$) can be also practically achieved in multi-resonant QM when the series of spectral-topological matching conditions for matching the spectroscopic parameters of interacting resonators are met \cite{Perminov2019Superefficient}.
Applying the idea of Flurin's work \cite{Flurin2015}, we have also shown \cite{Moiseev2020} the possibility of on demand QM on a system of four resonators with common resonator coupled with an external qubit (resonator) via controlled switch.
Perfect operation of this QM occurs when using the optimal parameters of the resonators ensure two conditions: A) - the fulfillment of impedance matching condition and B) - preserving the periodic structure of resonant frequencies in the multi-resonator system during transfer and storage stages of signal fields.

In this work, as a first step we develop efficient on demand multi-resonator QM coupled via controlled switch with external waveguide.
Such QM is necessary for transmitting quantum information over long distances, connecting to various quantum devices \cite{Wehner2018} and can operate in the optical frequency range.
The possibility of achieving very high efficiency and fidelity in the proposed multi-resonator QM with a switch is based on the fact that the switching of the connection of the common resonator with the waveguide occurs at times when all the stored signal radiation is in the mini-resonators. 
Due to this, the quantum noises arising in the common resonator caused by the active phase of the switch operation will be strongly fenced off from penetration into the microresonators, where the signal radiation is stored.

The difference between this QM and the previously studied scheme in \cite{Moiseev2020} is that the interaction with an external waveguide makes the considered resonator system an open (non-Hamiltonian) system where the interaction with the waveguide acquires a relaxation character which in another way affects the nature of spectral changes in the system of coupled resonators.
Therefore, the direct application of the results obtained in the work \cite{Moiseev2020} turns out to be impossible, as well as the solution that can ensure the preservation of both (A and B) conditions of the parameter matching becomes unobvious.
Further, we show that in the studied system of three resonators interacting with one common resonator connected to waveguide via a switch, it is possible to find the optimal parameters of this system for implementing  on demand efficient QM. 
Used methods and solution found are presented and the existing experimental possibilities of its implementation are discussed including  optical frequency range based on using the latest achievements in the fabrication of high-Q resonators and fast switches.

\section{Physical model}
\begin{figure}[t]
	\includegraphics[width = 0.45\textwidth]{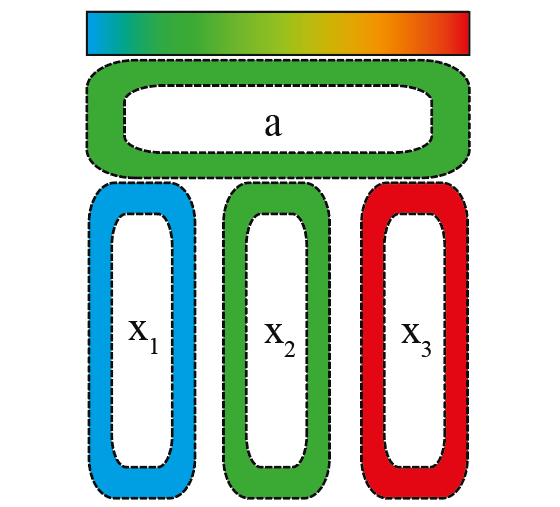}
	\caption{Principal scheme of a multiresonator quantum memory with switcher, which is integrated with an external waveguide.}
	\label{Scheme}
\end{figure}

Fig .~\ref{Scheme} shows a principal scheme of a multiresonator QM with three mini-resonators and one common resonator  connected to an external waveguide via a switch.
The mini-resonators have the following eigen frequencies: $\omega_n=\omega_0+\Delta_n$, where 
$\Delta_n=\{-\Delta,0,\Delta\}$ is the frequency offsets of the side mini-resonators from the common resonator with central frequency $\omega_0$, $f_n=f$ is the  coupling constants of the mini-resonators with the common resonator.
We assume that a switch can quickly change the coupling constant of the common resonator with the waveguide from $\kappa=0$ to constant $\kappa=\kappa_0$. 
Thus, we consider only two modes of operation when the common resonators has   these two values of the coupling constant $\kappa$.  

Taking into account the high quality factor of the resonators, we neglect the field attenuation in the description of the dynamics of the studied QM at times $t$, assuming $ t \ll 2 Q⁄\omega_0$
($n=\{1,...,N\}$) and, using frequency units, we set the Hamiltonian in the form:

\begin{eqnarray}
\nonumber H=\int d\omega \omega a_{\omega}^{\dagger} a_{\omega} +
{\sum}_{n} \omega_n b_n^{\dagger} b_n + \omega_0 a^{\dagger}a &\\+
\left[ {\sum}_{n} f_n a^{\dagger} b_n + g_{cw}\int d\omega a_{\omega}^{\dagger} a+h.c.\right],
\label{hamilt}
\end{eqnarray}
\noindent
here $a^{\dagger}$, $a$ and $b_n^{\dagger}$, $b_n$ are the  creation and annihilation Bose operators of the mode of the common resonator and the mode of the $n$th mini-resonator ($[a,a^{\dagger}]=1,[b_n,b_m^{\dagger}]=\delta_{n,m}$); $a_{\omega}^{\dagger}$, $a_{\omega}$ are the operators of creation and annihilation of the $\omega$-th mode of the waveguide ($[a_{\omega'},a_{\omega}^{\dagger}]=\delta(\omega'-\omega)$ ).

We describe the quantum dynamics using the well-known input-output formalism of quantum optics \cite{walls2007quantum} recently generalized to resonator-waveguide circuits \cite{Lalumiere2013} and well-recommended in the resonator microwave QMs \cite{Flurin2015,Matanin2022}.

We assume that all four resonators are prepared in the ground states $|G_r\rangle=\prod_{n=0}^3|g_n\rangle$ before signal pulses described by the initial state $|\psi_{in}\rangle$ is launched into the common resonator. 
Using Hamiltonian \eqref{hamilt} and following the input-output approach \cite{walls2007quantum} for the storage stage of an input signal field, we get the system of Langevin-Heisenberg equations for the resonator modes $a(t), b_{n}(t)$: 
\begin{eqnarray}
 \left[\partial_{t}+\frac{\gamma_n}{2}+i\Delta_n\right]b_{n}+if_n a &= \sqrt{\gamma} F_n, 
\nonumber \\
\left[\partial_t + \frac{k+\gamma_0}{2}\right] a+
i {\sum}_{n} f_n b_{n} & =
\sqrt{k} a_{in}+\sqrt{\gamma_0} F_0,
\label{eq_gen}
\end{eqnarray}

\noindent
where we have taken into account a relaxation of the cavity modes with decay constants $\gamma_n, \gamma_0$ and related Langevin Forces \cite{Scully1997}: $F_n(t), F_0 (t)$ ($[F_m (t),F_{m'}^{\dagger} (t')]=\delta k=2\pi g_{cw}^2$).

The input signal field $a_{in}(t)$ excites the common resonator, input- and outfields of the waveguide are coupled by the equation: $a_{in}(t)-a_{out}(t)=\sqrt{\kappa_0}a(t)$ \cite{walls2007quantum}:
($a_{in,out}(t)=\frac{1}{\sqrt{2\pi}}\int d\omega e^{-i\omega t} \tilde{a}_{in,out}(\omega)$).
The equations \eqref{eq_gen} are found for the Fourier components:
$(\tilde{b}_n(\omega'),\tilde{a}(\omega'))  =\frac{1}{\sqrt{2\pi}}\int d t e^{i\omega't} (b_n(t),a(t))$ and similar expressions for Langevin forces $\tilde{F}_{0}(\omega)$, $\tilde{F}_{n}(\omega)$ ($[\tilde{F}_{n}^{\dagger}(\omega'),\tilde{F}_{m}(\omega)]=\delta_{m,m'} \delta(\omega'-\omega)$):
\fla{
& \tilde{a} (\omega)=2\frac{\sqrt{\kappa_0}\tilde{a}_{\text{in}} (\omega)+\sqrt{\gamma_0}\tilde{F}_{\Sigma}(\omega)}{k+\gamma_0-2i\omega +2\chi (\omega)},
\label{a_solution}
}
\noindent
where 
$\chi(\omega)=Re \{\chi (\omega)\}+i\cdot Im \{\chi (\omega)\}$ 
is an effective permittivity of the memory with real and imaginary parts: 
\fla{
 Re \{ \chi (\omega)\} =\sum_{n=1}^{3}\frac{f_n^{2}\gamma_n/2}{(\gamma_n/2)^{2}+(\Delta_{n}-\omega)^{2}}, \\
 Im \{ \chi (\omega) \}=\sum_{n=1}^{3}\frac{f_n^{2}(\omega-\Delta_{n})}{(\gamma/2)^{2}+(\Delta_{n}-\omega)^{2}}.
}

The solution \eqref{a_solution} contains an effective Langevin operator $\hat{F}_{\Sigma}(\omega)$:
\fla{
\tilde{F}_{\Sigma}(\omega)=\tilde{F}_{0}(\omega)-i\sum^{N}_{n=1}\frac{g_n\sqrt{\gamma_n/\gamma_0}}{\gamma_n/2+i(\Delta_n-\omega)} \tilde{F}_{n}(\omega),
}
that is essential for determining noise of the QM device at finite temperatures, where $\langle \tilde{F}_{n}^{\dagger}(\omega),\tilde{F}_{n}(\omega)\rangle=\langle \tilde{F}_{0}^{\dagger}(\omega),\tilde{F}_{0}(\omega)\rangle \approx n_{\text{bath}}(\omega_0)$.
Using \eqref{a_solution} and \eqref{eq_gen} we get solution for the mini-resonator modes 
\fla{
\tilde{b}_n(\omega)=\frac{-if_n\tilde{a}(\omega)+\sqrt{\gamma} \tilde{F}_n(\omega)}{\gamma/2+i(\Delta_n-\omega)},
\label{b_n}
}
\noindent
and applying relation between input and output field we also find   
\begin{eqnarray}
& \tilde{a}_{out}(\omega)=S(\omega)\cdot \tilde{a}_{in}(\omega)+\tilde{b}_{\text{noise}}(\omega), 
\label{eq::in-out}
\\
& \tilde{b}_{\text{noise}}(\omega)=2\frac{\sqrt{k\gamma_0}\tilde{F}_{\Sigma}(\omega)}{k+\gamma_{0}+2\chi (\omega)-2i\omega},
\label{eq::b_nois}
\end{eqnarray}
\noindent
where $S(\omega)$ is a spectral transfer function of the memory:
\fla{
S(\omega)=\frac{k-\gamma_{0}-2\chi (\omega)+2i\omega}{k+\gamma_{0}+2\chi (\omega)-2i\omega},
}
$\tilde{b}_{\text{noise}}(\omega)$ is a noise component in the output signal caused by the interaction with the bath modes of all the resonators.
Equation \eqref{eq::b_nois} shows that the influence of intracavity quantum noise is highly suppressed in proportion to the factor $\frac{\gamma_0}{k}$ for $k\gg \gamma_0$.

The transfer function $S(\omega)$ is characterised by eigen frequencies of resonator system which are highly sensitive to the interaction between the resonators modes.
Recently \cite{Perminov2019Superefficient}, it was found that the interaction greatly changes the arrangement of the eigen frequencies $\omega '_n$.
The behavior of the eigen frequencies experiences a topological transition near the impedance matching condition, while the frequency arrangement should take the form of a periodic frequency structure to ensure high memory efficiency similar to photon echo on atomic frequency combs (AFC) \cite{Dubetskii_1985,Riedmatten2008, Afzelius2010,Akhmedzhanov2016}. 
Moreover high efficiency is also possible under more general conditions \cite{Moiseev2020} when the eigen frequencies $\omega'_n$ become a multiple of a certain frequency, for example  $\omega '_n = [-4,-1,1,4]\cdot \Delta'$,  arranged symmetrically relative to zero frequency offset, where  $\Delta'$ becomes a function of several parameters $\Delta'=\Delta'(\Delta,f,k)$.

Highly efficient quantum storage means the realization of an almost perfect  delay of the signal field for a given time interval $T$ ($T= \frac{2\pi}{\Delta'}$) so that  $S(\omega)\cong e^{i\omega T}$ in the operating frequency range with spectral efficiency $E(\omega)=|S(\omega)|^2\cong 1$.
This condition also means  minimization of the reflected field $\langle a_{out}(t)\rangle\cong 0$ and emptying the common resonator $a(t<T)\cong 0$. 
At the same time, the search for optimal parameters of a multi-resonant system should be carried out taking into account the behavior of the stored signal when the resonators are disconnected from the external waveguide. 
Assuming instantaneous disconnection of the resonators at a time $t=t_0$, we obtain the following system of equations for the resonator modes for $t>t_0$:
\fla{
 \partial_{t}b_{n}(t)  + if_n a(t) = &\left[\delta(t-t_0) -(\frac{\gamma}{2}+i\Delta_n)\right]  b_{n}(t)+
 \nonumber \\  
 & \sqrt{\gamma} F_n(t)\eta (t-t_0), 
\nonumber \\
\partial_t a(t) +
i {\sum}_{n}  f_n b_{n}(t) =&\left[\delta(t-t_0)-\frac{\gamma_0}{2}\right] a(t)  +
\nonumber \\
  &\sqrt{\gamma_0} F_0 (t) \eta(t-t_0).
\label{eq_gen-2}
}
\noindent
where $\eta(t-t_0)$ is a Heaviside function ($\eta(t-t_0)=1$ for $t>t_0$; and =0 for $t<t_0$).

In the general case, the equations \eqref{eq_gen-2} leads to the appearance of new eigen frequencies in the resonator system and these frequencies will not necessarily form a periodic structure of resonant lines necessary to ensure high efficiency of recovery of the stored signal. 

To increase the efficiency and operational functionality of the multi-resonator circuit, we apply methods of fast precision control and periodic positioning of the natural frequencies of coupled resonators together with impedance matching, using fast algorithms of algebraic and numerical optimal control \cite{Perminov2019Superefficient}.
Of great interest is the situation when the spectral width of the signal pulse becomes comparable with the spectral width of the comb of the natural frequencies of the resonators. Here we also show that the efficiency can be further improved by specializing the spectral profile of the field that carries quantum information between elements of a quantum device.

\section{Efficiency optimization procedure}
In order to achieve efficient QM in the proposed topology of resonators connection , it is required to optimize the controlled QM parameters in two operating modes, 
switching between which is carried out by controlling the connection of the common resonator with the external waveguide.
Assuming the presence of a periodic structure of the initial frequencies of three miniresonators (see fig. ~\ref{Scheme}), we start searching for the optimal coupling constant $f$ of these resonators with a common resonator that is disconnected from an external waveguide $k=0$ where the distribution of coupling constants is chosen in the form $[f_1,f_2,f_3]=f \cdot [0.8,1,0.8]$ for the possibility of obtaining an equidistant spectrum.
This situation corresponds to the long-term storage stage ($p=2$) of signal radiation in the multiresonator system, which requires a multiplicity of natural frequencies of the system, at which it becomes possible to periodically restore the stored quantum state in time.
For equal decay constants  ($\gamma_0=\gamma_n=\gamma$), we get the eigen frequencies  $\omega ^{(p)}_n=\tilde\omega^{(p)}_n-i\gamma$ of the multiresonator system ( $p=2$: $k=0$) where $\tilde\omega^{(p)}_n$ are the solutions of the following algebraic equation: 
\fla{
P(\tilde\omega)=\tilde\omega^4-(f_1^2+f_2^2+f_3^2+\Delta^2)\tilde\omega^2+\Delta^2f_2^2,\\
P(\tilde\omega)=0.
\label{condition-1}
}

\begin{figure}[ht]
	\includegraphics[width = 0.45\textwidth]{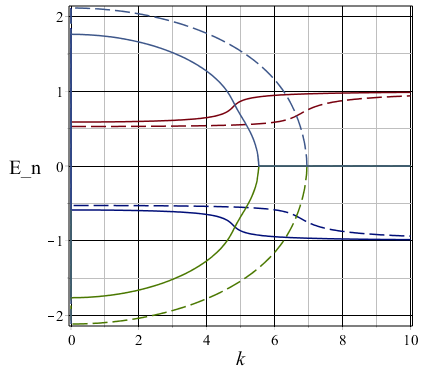}
	\caption{QM eigen frequencies $E_n=Re(\omega_n)$ for $f_n =1.119\cdot[1,1,1]$ -- dashed line, $f_n = 1.038\cdot[0.8,1,0.8]$ -- solid line.}
	\label{eigenmodes}
\end{figure}

The behavior of the eigen frequencies  $\tilde\omega^{(2)}_n$ depending on the constant coupling $f$ is shown in fig. \ref{eigenmodes}.
It is worth noting that with the growth of the coupling constant $k$ of the common resonator with an external waveguide, when certain critical values are reached ($k_1\approx 7$, $k_2\approx 5.5$), the two eigen frequencies merge into one and remain unchanged for each of these two cases considered (see fig. \ref{eigenmodes}).
This behavior characterizes a kind of topological transition found earlier for such systems in our work \cite{Perminov2019Superefficient}.
From the obtained numerical dependencies, we find  two sets of multiple eigen frequencies:  
$\tilde\omega_n^{(2)}(f_1)=[-4,-1,1,4]\cdot \Delta'(f_1)$ (where new spectral interval $\Delta'(f_1)=1.017$) at a coupling constant $f_1=1.119$) that reproduces results of work \cite{Moiseev2020}
and $\tilde\omega_n^{(2)}(f_2)=[-3,-1,1,3]\cdot \Delta'(f_2)$
($\Delta'(f_2)=0.588$ at a coupling constant $f_2=1.038$). 
The choice of such eigen frequencies makes it possible to readout the stored state at multiple intervals of time $T_n(f_1)=2\pi n/\Delta'(f_1)$ or $T_n(f_2)=2\pi n/\Delta'(f_2)$.

Assuming a given coupling constant $f_1$ or $f_2$, at the second step we can look for the optimal value of the constant coupling $k=\kappa(f_{n})$ of the common resonator with an external waveguide.
Similar to work \cite{Moiseev2018}, at the this stage of optimization, we impose a smoothness condition on the  phase delay $\tau(\omega)=-i Arg [S(\omega)]/ \omega$ at the point $\omega=0$, which can be represented as
$\partial^2_{\omega=0}[\tau]=0$,
where for near zero losses during the interaction with signal pulse temporal duration $\delta t$ ($\gamma_n \delta t\ll 1,\gamma_0 \delta t \ll 1$) in resonators transfer function $S_{11}(\omega)$:
\fla{
S_{11}(\omega)= 
-\frac{
k\omega(\Delta^2-\omega^2)-2i P(\omega)
}{
k\omega(\Delta^2-\omega^2)+2i P(\omega)
}, \\
\tau(\omega)= 
\frac{2}{\omega} \arctan \left[
\frac{k\omega(\Delta^2-\omega^2)}{2P(\omega)} \right].
\label{S-function}
}

\begin{figure}[t]
	\includegraphics[width = 0.45\textwidth]{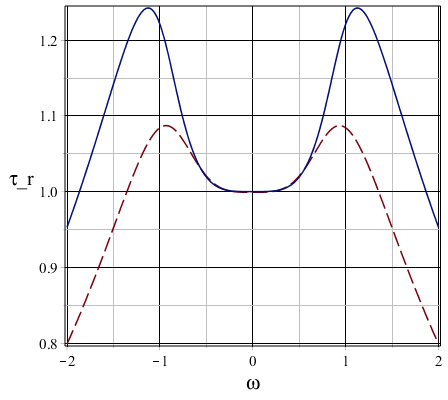}
	\caption{Relative phase delay $\tau_r=\tau(\omega)/\tau(0)$ for multiplicity spectra type $[-4,-1,1,4]$ -- dashed line, and equidistant spectra type $[-3,-1,1,3]$ -- solid line.}		
	\label{phase_delay}
\end{figure}

The transfer function $S_{11}(\omega)$ describes the spectral properties of the reflected signal. 
Strong suppression of the reflected signal in a certain spectral interval corresponds to the implementation of the impedance matching quantum memory \cite{Moiseev2010,Afzelius2010}.
From which we obtain the following smoothness condition for the coupling constants $k=\kappa$ and $f_n$ in algebraic-type form:
\fla{
\kappa=\sqrt{
\frac{\Delta^2f_2^2+(f_1^2+f_2^2+f_3^2)(f_1^2+f_3^2)}{\Delta^2/12}
}.
\label{condition}
}
This condition is necessary for the complete loading of the signal from the external waveguide into the QM.
For $f_n =1.119\cdot[1,1,1]$ and multiplicity spectra type $[-4,-1,1,4]$ we will get $\kappa_1=7.256$, for $f_n = 1.038\cdot[0.8,1,0.8]$ and equidistant spectra type $[-3,-1,1,3]$ we will get $\kappa_2=5.546$.
Figure \ref{phase_delay} shows a phase delay $\tau(\omega)$ with a smoothed region near the central frequency and which has a plateau corresponding to the operating spectral range of QM.

For the case of coupling ($p=1$: $k=\kappa$) \eqref{condition} in equation $S^{-1}(\omega^{(1)}_n)=0$, we find the dependence of 4 eigen frequencies $Re (\omega^{(1)}_n)$ from  the coupling constant $f$ and the similarly we find  the dependence $Re (\omega^{(1)}_n)$ from the constant $g$.
By assuming the equal decay constants  ($\gamma_0=\gamma_n=\gamma$), we get the eigen frequencies  $\omega ^{(p)}_n=\tilde\omega^{(p)}_n-i\gamma$ of the multiresonator system for the two cases ($p=1$: $k=\kappa$ and $p=2$: $k=0$) where $\tilde\omega^{(p)}_n$ are the solutions of the following algebraic equation: 
\fla{
2 f^2(\Delta^2-3 \tilde\omega^2) -\tilde\omega (\Delta^2 - \tilde\omega^2)(2 \tilde\omega + i k )=0,
\label{condition-2}
}
where we also assume that the coupling constant $k(t)$ changes fast enough between its two values $k=\kappa$ and $k=0$ that the evolution of radiation during the  switching time $\delta t_s$ when $dk/dt\neq 0$ can be neglected.

The frequencies $\omega^{(1)}_n$ determine the poles of the transfer function $S(\omega)$, describing the stationary absorption spectrum of a loaded multi-resonator system, while  $\omega^{(2)}_n$ are the eigen frequencies of a disconnected system where $Im[\omega^{(2)}_n]=0$ if $\gamma=0$.
In fig. \ref{eigenmodes} the eigen frequencies are presented for both cases ($p=1,2$), where we see that when the coupling constant $k(t)$ is changed between the two states ($k=\kappa$ and $k=0$), the energy spectrum and, accordingly, the internal dynamics of the multiresonator system changes significantly.
Thus, a natural problem is a matching the energy spectra of the system in different operating modes with on and off coupling, in which it will be possible to synchronize the dynamics of the two stages of evolution, allowing both the effective loading of signal pulse and its perfect on demand retrieval after long-term storage. In the proposed scheme, this problem is solvable due to the controlled flexible spectral characteristics of the QM with a common central resonator.

\section{Recording dynamics}
Due to the fact that there is a time  interval near $t=T_0/2$, when almost all the energy of the initial signal pulse is transfered in mini-resonators, and in the presence of a rapidly switching coupling $k(t)$, QM can be transferred at this moment of time into operation for long-term storage. 
In this case, the switching process can be noiseless due to the fact that there is almost no energy in the common resonator and waveguide. 
After several cycles in the mode of long-term storage of a signal pulse in a system of high-Q resonators, which has reversible temporal dynamics, we can similarly turn on the $k(t)$ coupling and on demand readout the signal pulse into an external waveguide.
\begin{figure}[ht]
	\includegraphics[width = 0.45\textwidth]{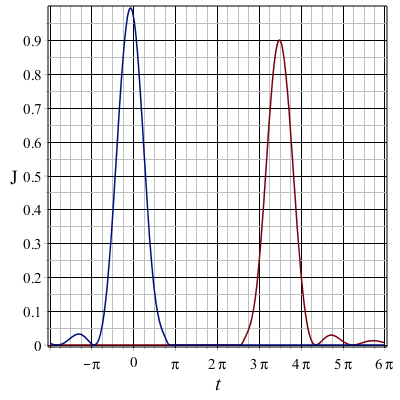}
	\caption{Relative intensity $J(t)$ of the echo signal for a Gaussian-type pulse with half-width $\sigma=1$.}
	\label{echo}
\end{figure}

\section{Conclusion}
We have demonstrated the ability to control two different regimes of efficient quantum storage in mutiresonator QM connected with waveguide via switchable coupling.
The optimal parameters of this quantum memory are found, at which it becomes possible to efficiently load signal pulses and store them for an arbitrary time, determined only by the intrinsic quality-factor of the resonators used.
It is noteworthy that the most optimal coupling constants of a multi-resonator quantum memory with an external waveguide are established near the region of the topological phase transition depending on the eigen frequencies of the magnitude of the coupling constants.
Interestingly, the presence of this type of spectral-topological phase transition is associated with the impedance matching condition and with the appearance of the possibility of implementing a periodic or multiple structure of eigen  frequencies in the multiresonator quantum memory, which is of great importance for the implementation of long-term on demand quantum storage.

The proposed mutiresonator QM can be implemented for  microwave fields in a system of high-Q coplanar superconducting resonators  where efficiency more than 60\% and fidelity near to 100\%  have been demonstrated for a microwave pulse attenuated to a single-photon level \cite{Matanin2022}.
The main technological advantage of the proposed QM circuit is its compactness due to a small number of miniresonators and the possibility of quick adjustment to the operating regime based on the use of specified optimal algebraic relations for the controlled parameters of the resonators.

The practical implementation of  the multiresonator QM in the optical range is also possible using recently appeared new high-performance components, such as microresonators with ultra-high Q-factor
\cite{Yang2016,Yang2018,Anderson:18,Gao:22}, which can be efficiently and accurately controlled \cite{Gao_2021,Li2020,Ahmed:19} and combined with resonant long-lived single atoms \cite{Moiseev2022} for a significant increase in the performance of quantum computing and due to long-term QM and optimization of geometric form factor quantum machines.

The considered quantum memory promises rich functionality both for the implementation of various dynamic storage modes and for the operation with the quantum information stored  due to the use of rich dynamics in the interacting resonators and the ability to readout information from different resonators at different times by controlled switcher \cite{Moiseev2020,Ahmed:19}.

\ack
The work was financially supported as part of the work on the "Integrated Quantum Optics" Lab KNRTU-KAI (basic idea, numerical simulation and analysis of results -- S.A.M., N.S.P.) and financially supported in the framework of the budgetary theme of Zavoisky Physical-Technical Institute (algebraic optimization methods for quantum systems -- N.S.P., S.A.M.).

\section*{References}

\bibliographystyle{iopart-num}
\bibliography{I_MR_QM}

\end{document}